\pdfoutput=1
\documentclass[acmsmall]{acmart}
\AtBeginDocument{%
  \providecommand\BibTeX{{%
    \normalfont B\kern-0.5em{\scshape i\kern-0.25em b}\kern-0.8em\TeX}}}
\usepackage{multirow}
\usepackage{xcolor} 
\usepackage{graphicx}
\usepackage{subcaption}
\newenvironment{renv}
  {\color{black}} 
  {} 

\newcommand{\italquote}[1]{\begin{quote}``\textit{#1}''\end{quote}}
\newcommand{\redacted}[1]{#1}
\newcommand{\topic}[1]{\textcolor{black}{#1}}
\newcommand{\revision}[1]{\textcolor{black}{#1}}

\setcopyright{acmcopyright}
\copyrightyear{2018}
\acmYear{2018}
\acmDOI{XXXXXXX.XXXXXXX}

\acmConference[Conference acronym 'XX]{Make sure to enter the correct
  conference title from your rights confirmation emai}{June 03--05,
  2018}{Woodstock, NY}
\acmBooktitle{Woodstock '18: ACM Symposium on Neural Gaze Detection,
 June 03--05, 2018, Woodstock, NY} 
\acmPrice{15.00}
\acmISBN{978-1-4503-XXXX-X/18/06}

\begin{document}

\title[Disruptors of Engagement with DMH Tools]{Investigating the Role of Situational Disruptors in Engagement with Digital Mental Health Tools}


\author{Ananya Bhattacharjee}
\affiliation{%
  \institution{Computer Science, University of Toronto}
  \city{Toronto}
  \state{Ontario}
  \country{Canada}
}

\author{Joseph Jay Williams}
\affiliation{%
  \institution{Computer Science, University of Toronto}
  \city{Toronto}
  \state{Ontario}
  \country{Canada}
}

\author{Miranda L. Beltzer}
\affiliation{%
  \institution{Preventive Medicine, Northwestern University}
  \city{Chicago}
  \state{Illinois}
  \country{USA}
}

\author{Jonah Meyerhoff}
\affiliation{%
  \institution{Preventive Medicine, Northwestern University}
  \city{Chicago}
  \state{Illinois}
  \country{USA}
}

\author{Harsh Kumar}
\affiliation{%
  \institution{Computer Science, University of Toronto}
  \city{Toronto}
  \state{Ontario}
  \country{Canada}
}

\author{Haochen Song}
\affiliation{%
  \institution{Statistical Sciences, University of Toronto}
  \city{Toronto}
  \state{Ontario}
  \country{Canada}
}

\author{David C. Mohr}
\affiliation{%
  \institution{Preventive Medicine, Northwestern University}
  \city{Chicago}
  \state{Illinois}
  \country{USA}
}

\author{Alex Mariakakis}
\affiliation{%
  \institution{Computer Science, University of Toronto}
  \city{Toronto}
  \state{Ontario}
  \country{Canada}
}

\author{Rachel Kornfield}
\affiliation{%
  \institution{Preventive Medicine, Northwestern University}
  \city{Chicago}
  \state{Illinois}
  \country{USA}
}

\renewcommand{\shortauthors}{Bhattacharjee et al.}

\begin{abstract}
Challenges in engagement with digital mental health (DMH) tools are commonly addressed through technical enhancements and algorithmic interventions. This paper shifts the focus towards the role of users' broader social context as a significant factor in engagement. Through an eight-week text messaging program aimed at enhancing psychological wellbeing, we recruited 20 participants to help us identify situational engagement disruptors (SEDs), including personal responsibilities, professional obligations, and unexpected health issues. In follow-up design workshops with 25 participants, we explored potential solutions that address such SEDs: prioritizing self-care through structured goal-setting, alternative framings for disengagement, and utilization of external resources. Our findings challenge conventional perspectives on engagement and offer actionable design implications for future DMH tools.

\end{abstract}




\maketitle
\section{Introduction}

Mental health conditions such as anxiety and depression are highly prevalent among adults \cite{vahratian2021symptoms, thomeer2023racial}. These issues are often associated with substantial distress and disability and can lead to serious disruptions in areas like work, education, and social relationships \cite{goodwin2022trends, kessler2022estimated}. Traditional therapeutic interventions can be effective in alleviating mental health symptoms and improving overall wellbeing, but are often inaccessible due to barriers like cost, societal stigma, or a preference for self-reliance \cite{gulliver2010perceived, robards2018marginalized, oguamanam2023intersectional, lattie2020designing, frik2023model}. 
These barriers have led to the development of digital mental health (DMH) tools accessible through everyday devices like computers and smartphones to complement traditional interventions \cite{fairburn2017impact, mohr2017personal, bhattacharjee2023investigating}. Their versatility has been showcased across a spectrum of settings, offering notable benefits like accessibility and personalization of care \cite{lattie2022overview, kornfield2022meeting, pendse2022treatment, mohr2018solution}.
However, frequent user disengagement presents a significant obstacle to the effective utilization of DMH tools \cite{borghouts2021barriers}. 

In the context of self-guided DMH tools, engagement broadly refers to the process through which a user creates, sustains, and eventually terminates a relationship with a computerized system over time \cite{sidner2004look, xu2013designing, doherty2018engagement}. Although it is not necessary for every DMH tool to be continuously used over weeks or months, the acquisition of knowledge and self-management skills through many such tools is predicated on their usage over an extended period \cite{scholten2019use, pendse2022treatment}. Thus, engagement often focuses on the extent to which an individual makes productive use of a tool in their daily life \cite{schueller2017integrating, zech2023integrative}. While many individuals are eager to try out new DMH tools, they often lack the sustained engagement needed to support learning and applying new skills and perspectives \cite{muench2017more, lipschitz2023engagement}. As such, the benefits of DMH tools that require prolonged use often remain unrealized relative to their potential \cite{figueroa2022daily, bhattacharjee2023integrating}. 

Several studies have explored the underlying causes of disengagement with DMH tools, identifying factors such as the provision of generic content and failures to recognize the optimal moments for user intervention \cite{bhattacharjee2023investigating, rennick2016health, bhattacharjee2022design, kornfield2020energy}.
To address these issues, HCI and CSCW researchers have largely concentrated on integrating new DMH tool features that enhance user interest (e.g., coaching, peer-to-peer communication, gamification), improve usability, and tailor content or timing to users' needs \cite{saleem2021understanding, dwivedi2019re, borghouts2021barriers, kornfield2022involving, ng2019provider, yoo2024missed}. However, recent studies have also noted the limits of such 
approaches to improving engagement when disengagement can also stem from fundamental disruptions in daily life that dramatically shift user's priorities~\cite{lipschitz2023engagement, pendse2022treatment}. For instance, an unexpected change in one's work schedule or a sudden financial crisis might derail their ability or motivation to engage with a DMH tool, regardless of how well the tool is designed or tailored to their preferences. These circumstantial issues are tied to social context, which encompasses the specific settings in which human relationships and personal circumstances influence the way one thinks and behaves~\cite{dourish2004we, bronfenbrenner1977toward, helliwell2004social}. The impact of social context on engagement is well-established in face-to-face therapy, where issues like job constraints and financial hardship are often cited as reasons for early termination, sometimes more so than dissatisfaction with therapy  \cite{renk2002reasons, roe2006clients, hynan1990client}. Since DMH tools are intended to be more convenient and self-directed, the role of circumstantial factors in disrupting user engagement with them is less studied.

Given that key factors related to one's social context can profoundly influence individuals' goals and priorities \cite{gallie2019research, allen2014social, slavich2020social, tachtler2020supporting, sabie2020memory, bhattacharjee2022kind}, \revision{we are interested in exploring how these factors impact their use of DMH tools that require ongoing engagement over long periods during which one's social context is likely to evolve.} Changes in social context can exacerbate mental health symptoms, inducing a sense of overwhelm and undermining the prioritization of self-care \cite{slavich2020social, mills2020prioritising}; these setbacks pose a fundamental challenge to sustained usage of these tools.
We specifically focus on text-messaging DMH tools designed for long-term use spanning weeks to months. 
Text messaging is quickly becoming one of the well-established DMH platforms for promoting psychological wellbeing \cite{bhattacharjee2023investigating,  kornfield2022meeting}. Compared to alternatives like mobile applications and online programs, text messaging has been embraced for its wider reach and accessibility in supporting psychological wellbeing \cite{kretzschmar2019can, chikersal2020understanding, rathbone2017use, inkster2018empathy, morris2018towards}.
In addition, reflecting their long-term use, this focus allows for insight into engagement patterns as they emerge over time and across changes in life circumstances.

These considerations motivate the following two research questions:

\begin{itemize}
\item \textbf{RQ1:} How does an individual's social context influence their engagement with DMH tools \revision{that require extended use spanning weeks or months}?
\item \textbf{RQ2:} How can \revision{those} DMH tools be designed to better account for the challenges and opportunities presented by an individual's unique social context?
\end{itemize}

\noindent
By investigating these questions and developing a deeper understanding of how one's broad societal context shapes engagement, designers may be able to create tools that enable users to navigate challenges without disrupting their interaction with the system. 
Furthermore, DMH tools may be designed in such a way that they are not just resilient in the face of these disruptors but are actively helpful in navigating them. This shift in perspective may allow the DMH tools to be supportive of users' unique situations, rather than merely functioning in spite of them. DMH tools that require shorter adherence periods (e.g., a few minutes or hours) may not need to deeply consider how fluctuations in social context affect engagement and, hence, fall outside our study's scope.

We commenced our research by deploying an eight-week text messaging program for managing psychological wellbeing to 20 individuals. Utilizing two rounds of interviews, we examined the influence of participants' social contexts on their engagement with the program. Through this exploration, we identified various situational engagement disruptors (SEDs) that include academic challenges, workplace demands, family obligations, and unexpected life events such as health issues and relocation. \topic{These SEDs were often characterized by a sense of overwhelm -- the feeling of being unable to keep up with the various demands on oneself, marked by high levels of stress and a perceived loss of control over one’s life \cite{kabigting2019conceptual, hopps1995power}. Consequently, SEDs hindered participants' ability to prioritize self-care and dedicate time to interact with DMH tools.}
To address these disruptors, we conducted a subsequent study featuring five design workshops with 25 participants to identify strategies for incorporating these considerations into DMH tool design. Our research yielded several key design considerations: prioritizing self-care through structured goal-setting, designs that accommodate flexible engagement, and addressing SEDs within social contexts.
Our contributions include:

\begin{itemize}
    \item The inclusion of social context as a critical variable in the study of engagement with DMH tools,
    \item The identification of specific SEDs that impact user engagement, and
    \item The provision of targeted design strategies to mitigate the identified SEDs and enhance user engagement.
\end{itemize}

\section{Related Work}

In our overview of related work, we first outline the diverse social factors that negatively impact psychological wellbeing.
We then describe how DMH tools, more specifically text messaging services, have been used to promote behavior change and psychological wellbeing.
We follow this overview with a discussion on the challenges that designers and users face in sustaining engagement with DMH tools.

\subsection{Social Context and Individual Challenges as Determinants of Psychological Wellbeing}

Several frameworks have been developed in recent years to explain the role of social context and individual challenges as determinants of psychological wellbeing \cite{slavich2020social, slavich2023social, allen2014social, glanz2008health, tachtler2021unaccompanied, oguamanam2023intersectional, sallis2015ecological, yoo2024missed}. Among these, the Social Safety Theory \cite{slavich2020social} provides insights into the dichotomy of (1) socially safe environments characterized by elements such as acceptance, stability, and harmony and (2) socially threatening situations marked by elements such as conflict, turbulence, and unpredictability. According to the theory, socially threatening situations can exert a considerable impact on one's psychological wellbeing. Examples of this can be found in everyday situations like performing poorly on an exam or losing a job. These events can extend beyond their immediate academic or economic effects, leading to negative social evaluations, a decline in self-esteem, and a shift in social status \cite{diamond2022rethinking, noronha2018study, bhattacharjee2021understanding, katz2017self}. Adding to this understanding, one study \cite{allen2014social} delved into the complex interplay between living and working conditions, socio-economic standing, and the surrounding environment. Through a multi-level framework that encompassed various life stages and contextual factors, they illuminated the detrimental mental health consequences stemming from adverse situations such as unemployment, inadequate education, and instances of gender discrimination.

Many studies have investigated the connections between social factors and psychological wellbeing across diverse settings and demographic groups \cite{brydsten2018health, lecerof2015does, han2015social, reibling2017depressed, li2022suffered, groot2022impact, de2017gender, frik2023model, oguamanam2023intersectional, murnane2018personal, jung2023enjoy}. For example, economic challenges have been found to significantly impact psychological wellbeing. One study \cite{guan2022financial} discovered that tangible possessions and financial strain were strong predictors of depressive symptoms, while another \cite{katz2017self} found a correlation between lower income and symptoms of depression among transgender individuals in the USA. Among university students, academic performance, parental expectations, and economic stress have been identified as primary stressors \cite{de2016relationship, bruffaerts2018mental, lattie2020designing}. Psychological wellbeing was also found to be influenced by factors such as social isolation and physical health issues in older adults \cite{carod2017social, mccrone2008paying}, deportation fears among immigrant youths \cite{tachtler2021unaccompanied, tachtler2020supporting}, and pandemic-related stress in parents \cite{li2022suffered}.



Recognizing the influence of various socio-economic challenges on mental health outcomes, some studies developed targeted interventions that address the unique challenges faced by specific groups. These groups include but are not limited to low-income populations, minoritized populations, and those at risk of homelessness \cite{aubry2016multiple, kerman2018effects, leung2015household, murnane2018personal, sallis2015ecological, li2010stage, schueller2019use}. Interventions in this space are structured to account for the technological resources available to these targeted populations and the underlying issues contributing to their mental health conditions. 
For instance, targeted efforts to improve housing conditions have been shown to lessen the demand for inpatient psychiatric services among individuals who are unhoused and suffer from mental disorders \cite{kerman2018effects}. 
Other studies advocate for considering temporal aspects like daily routines and life transitions when designing interventions for psychological wellbeing \cite{doherty2010fieldwork, murnane2018personal, reddy2006temporality, jung2023enjoy, bhattacharjee2023investigating}. Implementation science literature \cite{graham2020implementation, urquhart2020defining, powell2015refined} complements these works by emphasizing the importance of social context and structures that can either facilitate or impede the effective use of mental health resources. Rather than concentrating solely on tool design, this field explores strategies for successfully deploying existing tools in specific settings. Such strategies often include user and staff training as well as monitoring tool uptake \cite{powell2015refined, powell2012compilation}. Collectively, these works underscore the importance of a holistic approach that integrates mental wellbeing considerations into broader societal contexts.

\subsection{Text Messaging as a DMH Platform for Promoting
Behavior Change and Psychological Wellbeing}
Text messaging has emerged as a powerful DMH platform, providing an accessible and personalized means of supporting individuals in their health journeys. The broad reach of this medium extends across various demographic groups, making it particularly useful for populations that encounter obstacles in accessing more conventional digital platforms like apps and online programs \cite{muench2017more, bhattacharjee2022kind}.
In recent years, text messaging interventions have shown considerable success in supporting behavior change across various physical and mental health challenges \cite{haug2013efficacy, haug2013pre, yun2013text, suffoletto2023effectiveness, shalaby2022text, figueroa2022daily, bhandari2022effectiveness, villanti2022tailored, gipson2019effects}. For instance, \citet{villanti2022tailored} conducted a randomized controlled trial and found that a tailored text message intervention for socioeconomically disadvantaged smokers led to greater smoking abstinence rates and increased desire to cease smoking compared to online resources. A similar study \cite{liao2018effectiveness} demonstrated that frequent text messages (3-5 messages per day) could contribute to lower cigarette consumption rates among adult smokers. 
Regarding other forms of behavior change, \citet{gipson2019effects} developed an intervention to improve sleep hygiene among college students, and \citet{glasner2022promising} developed an intervention to improve medication adherence and reduce heavy drinking in adults with HIV and substance use disorders. 
Other areas where text messaging has shown promise include weight management \cite{siopis2015systematic, donaldson2014text}, promoting physical activity \cite{kim2013text, smith2020text, murnane2020designing}, and enhancing vaccine uptake \cite{buttenheim2022effects, mehta2022effect}, among many others.

There has also been a surge of text-messaging services aimed at promoting psychological wellbeing~\cite{kretzschmar2019can, chikersal2020understanding, rathbone2017use, inkster2018empathy, morris2018towards, stowell2018designing, kornfield2022meeting, agyapong2020changes}. The breadth of the content they cover and the goals they seek to support are vast. Most text messaging services employ therapeutic techniques from clinical psychology — cognitive behavioral therapy \cite{willson2019cognitive}, dialectical behavior therapy \cite{linehan2014dbt}, acceptance and commitment therapy \cite{hayes2004acceptance}, and motivational interviewing \cite{hettema2005motivational} — to deliver support. Text messages can function as reminders, assisting individuals in setting aside time for leisure and physical exercise \cite{bhattacharjee2023investigating, figueroa2022daily}. They can also facilitate expressions of gratitude, nudging individuals to reflect on life's positive aspects \cite{bhattacharjee2022design}. Furthermore, text messages can share narratives that offer insights into how peers have navigated and surmounted challenging life situations, thereby aiding individuals in applying psychological strategies in their own lives \cite{bhattacharjee2022kind}. 

Several studies have leveraged the utility of text messaging as a medium to offer support for individuals grappling with various mental health challenges \cite{agyapong2020changes, levin2019outcomes, arps2018promoting, nobles2018identification, bhattacharjee2023investigating, agyapong2022text4hope}. \citet{agyapong2022text4hope} implemented such a system during the COVID-19 pandemic, dispatching daily support messages that led to decreased self-reported anxiety and stress. \citet{levin2019outcomes} combined psychoeducational content and reminders to assist those with bipolar disorder and hypertension. Meanwhile, \citet{arps2018promoting} tackled adolescent depression by sending daily texts centered around gratitude. Yet, there are several challenges pertaining to one's individual and social context that could prevent text messaging systems and other DMH tools from providing more benefits to users. We discuss these issues in the next subsection.

\subsection{Challenges of Sustaining Engagement with DMH Tools}

\revision{Prior work has delineated engagement through two conceptual constructs: as a subjective experience and as a behavioral phenomenon \cite{zech2023integrative, perski2017conceptualising}. Characterized by focused attention, invested time, intrinsic motivation, and enjoyment during interactions with a system \cite{mihaly1990flow, o2008user, brown2004grounded}, the subjective aspect of engagement is akin to the psychological states of flow and immersion. Meanwhile, behavioral science literature defines engagement as the nature of an individual's interaction with a DMH tool according to measures such as usage frequency, duration, depth, and adherence over time \cite{danaher2006defining, mcclure2013effect, mcclure2013effect}. Behavioral interpretations of engagement also distinguish between active participation (e.g., contributing content through posting) and passive involvement (e.g., reading without commenting) \cite{bhattacharjee2023investigating, burke2010social, yardley2015person}, suggesting that engagement can be seen as a dynamic process that includes varying levels of user involvement and interaction patterns over time \cite{o2008user}.}

\revision{Regardless of the specific definitions and metrics employed to assess engagement, studies often report low engagement rates with DMH tools \cite{baumel2019objective, lipschitz2023engagement}. \citet{waller2009barriers} found that clients may be twice as likely to discontinue the use of a DMH intervention compared to other forms of therapy.} \citet{baumel2019objective} reported that individuals with a mental health app installed exhibit a median daily open rate of 4.0\%, and the 15-day median retention rate stands at 3.9\%. These statistics suggest that users might access DMH tools sporadically, posing a challenge for interventions that require sustained engagement over a prolonged period.


Prior work has identified some elements of DMH tools that can cause disengagement. A common issue is the use of generic and repetitive content \cite{bhattacharjee2022design, rennick2016health, brown2014mobile}. While aiming for a wide reach, DMH interventions risk seeming insincere as generalized material may not resonate with users' unique situations \cite{bhattacharjee2023investigating, slovak2023designing, jardine2023between}. This lack of personal relevance, compounded by limited content variety, often leads to decreased engagement due to monotony and boredom \cite{brown2014mobile, jardine2023between}.
Furthermore, DMH tools face the challenge of competing with various ongoing and urgent demands for user attention \cite{muench2017more}. Even with personalized content, these tools run the risk of being ignored if they do not intervene at opportune moments \cite{bhattacharjee2022design}.

Researchers have generally endeavored to address these barriers and enhance engagement with DMH tools by refining features that inhibit engagement, making improvements in UI/UX, and tailoring algorithms to ensure content and delivery personalization  \cite{doherty2012engagement, harding2015hci, poole2013hci, xu2013designing, howe2022design, lipschitz2023engagement, yardley2016understanding}. This view toward engagement is also prominent in the design of influential technology adoption models, such as the Technology Acceptance Model (TAM) \cite{venkatesh2008technology, davis1989perceived} and the Unified Theory of Acceptance and Use of Technology (UTAUT) \cite{venkatesh2003user, dwivedi2019re}. 
These models have centered on factors like perceived usefulness and ease of use. 
Social influence has been included but to a lesser extent, with an emphasis on elements such as peer influence and subjective norms \cite{marangunic2015technology}. Some initiatives have proactively identified individuals at risk of low engagement to implement adaptive interventions, such as providing specialized coaching or peer support to enhance their engagement \cite{arnold2019predicting, lipschitz2023engagement, jung2023enjoy}. 

\revision{While these efforts mark significant strides in enhancing DMH tool engagement, research on how changes in an individual's social context impact engagement dynamics is still limited. \citet{ritterband2009behavior} discussed the potential impact that external factors like family and professional responsibilities might have on engagement with interventions, advocating for more in-depth research. Financial conditions, material resources, and time constraints have also been hypothesized to significantly influence engagement \cite{short2015designing,  perski2017conceptualising, yardley2016understanding, kaziunas2017caring}. \citet{borghouts2021barriers} emphasize that people's ability to fit DMH tools into their daily routines and socio-technical factors like intervention usability, data security, peer perception, and accessibility affect user engagement.}

\revision{Building on this prior work, we investigate how fluctuations in social context might impede the consistent and sustained engagement with DMH tools through the deployment of an eight-week-long text messaging program.}
These disruptors of engagement are likely to be multifaceted, unpredictable, and possibly challenging to monitor \cite{bhattacharjee2022design, allen2014social, slavich2023social, kaziunas2017caring}. They can manifest broadly and suddenly, leading to unexpected disruptions or even causing individuals to halt engagement altogether. Traditional user studies spanning only a few days or weeks may fail to capture these disruptors \cite{bhattacharjee2022design}. Instead, the influence of these factors may only emerge over time as users with mental health concerns integrate DMH tools into their daily lives. 
In our first study, we aim to bridge this gap by extending our understanding of how social contexts impact engagement with DMH tools. Through an eight-week-long study, we seek to identify these disruptors and provide a more comprehensive view of how these complex factors influence engagement. In our second study, we build on this knowledge by eliciting potential strategies for engaging end-users of DMH tools facing disruptors. 

\section{Procedure of Study 1}
We conducted our first study with 20 participants in an eight-week text messaging program designed to improve psychological wellbeing. The duration of the program enabled us to examine broader social contexts that hindered and shaped their engagement with a DMH tool. The study received approval from the Research Ethics Board at \redacted{Northwestern University}, and its logistics are described below.

\subsection{Participants}

Participant recruitment for this study was facilitated by \redacted{our research center's} Online Research Registry as well as a web-based screening platform of a nonprofit mental health advocacy organization called \redacted{Mental Health America (MHA)}. 
The study was presented as a field trial to test a text-messaging tool constructed to aid individuals in managing their symptoms of depression and anxiety. Across both recruitment sources, participants were required to be at least 18 years old, own a mobile phone, be a resident of the United States, and not currently receiving psychotherapy or planning to start psychotherapy in the study timeframe. 

Participants also completed self-screening surveys for depression and anxiety to confirm that they reflect potential users who stand to benefit from engaging with a DMH tool. Eligibility was determined based on self-reported scores from the Patient Health Questionnaire-9 (PHQ-9) \cite{kroenke2001phq} and the General Anxiety Disorder-7 (GAD-7) \cite{williams2014gad}. Those who scored 10 or higher on either questionnaire were considered eligible for participation.


The final cohort comprised 20 participants with an average age of $31.0\pm3.0$ years. The participants identified with multiple genders (17 women, 2 men, 1 non-binary) and several racial groups (10 White, 1 Black/African American, 1 Asian, 1 American Indian/Alaskan Native, 3 mixed race, and 4 undisclosed). 
\topic{Participants were asked whether a specialist or healthcare provider had ever diagnosed them with major depression or anxiety disorder; 11 participants (55.0\%) reported being diagnosed with depression, and 8 participants (40.0\%) reported being diagnosed with an anxiety disorder. However, a formal diagnosis of depression or anxiety was not required for participation.} We refer to these participants as FP1--FP20.

\subsection{Overview of the Text Messaging Program}

After undergoing the eligibility and consent processes, participants were enrolled in the Small Steps SMS program \cite{meyerhoff2024small} -- an automated, interactive program created by \citet{meyerhoff2024small} to build users' capability to self-manage depression and anxiety symptoms. \revision{The program sought to sustain engagement by offering a variety of ways for users to interact \cite{meyerhoff2024small}.} It offered daily interactive dialogues that taught users how to apply 11 evidence-based psychological strategies. The presentation of these strategies largely drew from works by \citet{kornfield2022involving} and \citet{meyerhoff2022system} 
and were grounded in theories including acceptance and commitment therapy \cite{hayes2006acceptance}, cognitive behavioral therapy \cite{willson2019cognitive}, and social rhythm therapy \cite{frank2022interpersonal}.

\begin{figure}
    \centering
    \includegraphics[width=1\linewidth]{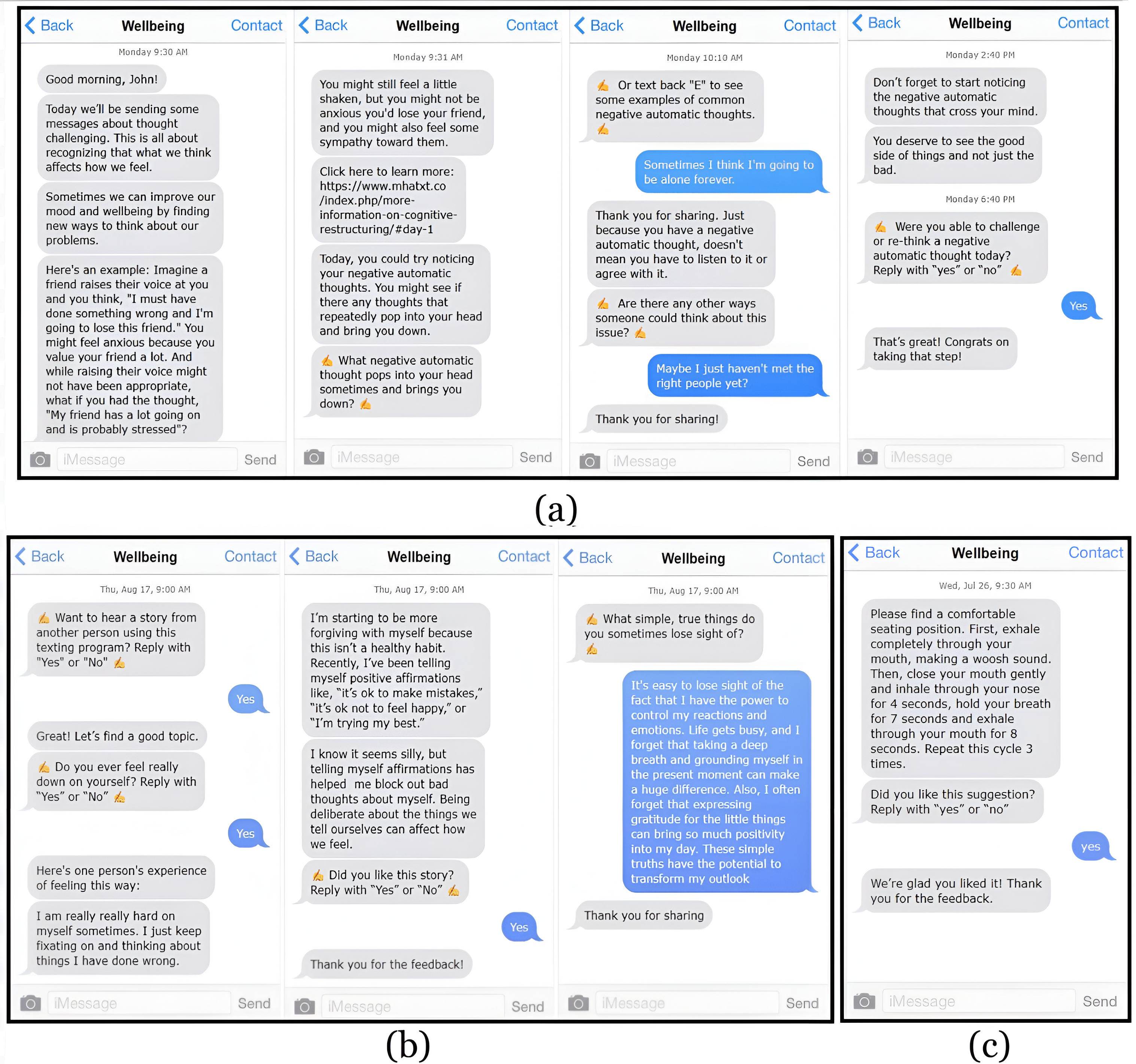}
    \caption{Examples of message dialogues seen by participants in the Small Steps SMS Program: (a) A dialogue showcasing specific psychological strategies and skill-building exercises, (b) illustrative accounts of individuals successfully employing strategies to overcome challenges, and (c) messages designed to prompt self-reflection.}
    \label{fig:conversations}
\end{figure}

The program provided active personalization, enabling users to select which psychological strategies they wanted to use over time. Within a subset of dialogues, users had the option to choose between different topics or exercises.
Figure \ref{fig:conversations} illustrates a subset of the message dialogues sent to users. Within each dialogue, users responded to sequential messages that used branching logic to tailor replies from the program. The program's daily dialogues varied over time in their approach.
Some of the messages included:
\begin{itemize}
    \item Introductions to specific psychological strategies and prompts to participate in skill-building exercises \cite{meyerhoff2022system},
    \item Stories of individuals who had successfully harnessed these strategies to navigate challenges \cite{bhattacharjee2022kind},
    \item Invitations to compose and receive supportive texts for peers \cite{meyerhoff2022system},
    \item Succinct tips or prompts for self-reflection, including messages that directed users to more extensive psychoeducational resources \cite{bhattacharjee2023investigating}, and 
    \item Prompts for users to think about or respond to open-ended questions \cite{bhattacharjee2023investigating} 
\end{itemize}
A more thorough description of the Small Steps SMS program can be found in \citet{meyerhoff2024small}, \topic{and additional diagrams illustrating the sequence of certain text messaging dialogues are provided in Appendix~\ref{sec: flow}.} \topic{The program spanned eight weeks, which aligns with typical durations for clinical DMH interventions and is comparable to the expected engagement period required for people to benefit from in-person therapy~\cite{lipschitz2022digital, mohr2019randomized}. 
This duration reflects that building new routines and incorporating new skills into daily life requires time, practice, and reinforcement. The Small Steps SMS program was specifically designed to provide opportunities not only for learning but also for implementing and reinforcing new habits in order to achieve downstream effects on symptoms. However, it is important to recognize that not all DMH tools require such extended durations; for example, some are designed as single-session interventions lasting around 10 to 20 minutes, focusing on teaching a specific micro-skill or providing immediate support for stress management  \cite{bhattacharjee2024exploring, paredes2014poptherapy, schleider2020future}}.

\subsection{Data Collection and Procedures}

Participants were asked to complete two semi-structured interviews, one midway and another at the end of the eight-week program, conducted via Zoom by two research team members. These interviews were intended to gather insights into how participants' social contexts, personal responsibilities, and other circumstances evolved over the course of the study and how these contexts influenced participants' interaction with the text messaging program. Examples of interview questions included:

\begin{itemize}
\item Have there been any changes in your interaction with or perceptions of the program over the last month? Can you describe these changes?
\item Over the past month, have there been any life changes that have influenced your usage of the program? Can you describe how they affected your use of the texting program?
\item Engagement levels can fluctuate in programs like this. What are your thoughts on why this might occur?
\item Have you observed any changes in your daily activities, personal life, or emotional state since starting the program? What were they?
\end{itemize}

All 20 program participants were invited to interviews, with all but one completing at least one. Eighteen participated in the mid-program interviews at four weeks, and 17 in the final interviews at eight weeks. Interview durations ranged from 20 to 45 minutes, with participants compensated \$20 USD per hour.

\subsection{Data Analysis}
After transcribing the interviews, our team undertook a thematic analysis \cite{cooper2012apa} of the qualitative data. Two team members, termed ``coders,'' began by thoroughly reviewing all transcripts to acquaint themselves with the content. Following an open-coding process \cite{khandkar2009open}, each coder independently created a preliminary codebook. Subsequent meetings enabled the team to consolidate the codes into a unified codebook. \topic{During these discussions, the team refined code definitions, identified recurring codes, assessed each code’s relevance to the research questions, and eliminated those that did not directly pertain to the research questions at hand.} The coders then tested the shared codebook on a subset of the data consisting of eight interview transcripts. This iterative process allowed for further refinements to the codebook. Once a consensus was reached, each coder applied the finalized codebook to separate halves of the remaining data.

\subsection{Ethical Considerations}
\label{study1_ethics}
The research team was comprised of faculty members and graduate students from various fields including computer science, human-computer interaction, cognitive science, and clinical psychology. Given that  mental health research presents multiple ethical challenges, we took measures to address these at all stages of the research.

Participants were explicitly informed at the commencement of the study that the text messaging system was not a crisis intervention service. They were also given a list of emergency contacts, including crisis text lines and suicide helplines, in the possible event that such information would be useful. We did not solicit suicide-related information at any point in the study; however, given the open-ended nature of text messaging, we recognized the unlikely possibility that participants may disclose unprompted suicidal thoughts or behaviors. Hence, we took measures to ensure the safety of participants. We reviewed text messages from participants on a daily basis. We were prepared to reach out to any participants indicating a risk of suicidal ideation or self-harm and conduct the Columbia Suicide Risk Assessment protocol \cite{posner2008columbia}. However, no such risks emerged during the study, and no follow-up assessment was necessary. 

Similar considerations were applied to the interviews, where interviewees were informed they could skip questions or exit at any time. Interviewers were trained to conduct the Columbia Suicide Risk Assessment protocol as well, but again, no such measures were necessary. 
\section{Findings from Study 1}

\topic{Participants offered varied interpretations of what constitutes a DMH tool. They reported using tools like Woebot (FP11) and mood-tracking apps (FP20) as well as leveraging social media for peer support (FP17). Others (FP4, FP5, and FP9) mentioned exploring digital reflection techniques online, while FP10 highlighted using apps for weight management. Several participants (FP5, FP17, FP19) had also mentioned that they had familiarity with some of the psychological concepts and techniques introduced by the Small Steps SMS program.} 

Our interviews illuminated various factors that disrupted user engagement with the Small Steps SMS program. While some experiences of disengagement were linked to program-specific issues like habituation and individual preferences, the majority stemmed from the broader social contexts affecting participants. We first briefly present disengagement factors linked to habituation and individual preferences before delving into those influenced by social context.

\subsection{Disengagement Due to Habituation and Personal Preferences}

During the eight-week program, participants demonstrated a relatively high level of responsiveness to the messages sent by the system. On average, they responded at least once on 70\% of the study days, with individual daily responsiveness spanning from 34\% to 100\%. The level of participation also declined over the study, with average daily responsiveness falling from 93\% in the first week, to 47\% in the 8th week. Participants relayed that these trends partly reflected diminishing novelty, as well as occasional dissatisfaction regarding the content and frequency of the messages.

\topic{A prominent reason for gradually decreasing responsiveness to the messages was habituation. Over the course of the program, participants became accustomed to the repetitive nature of the daily text messages, rendering them increasingly predictable. This predictability sometimes reduced the novelty of the messages and engendered a sense of monotony.} As the routine of reading and responding to messages became more ingrained, participants found the task increasingly tedious, eroding both their motivation to engage with the program and their perception of its utility. Over time, the diminishing returns of the repetitive interactions led participants to question the program's usefulness, thereby lowering their overall responsiveness. FP10 expressed this sentiment by saying:


\italquote{It’s like a daily thing that you constantly have to check up every day. Then, yeah, it’s like a chore. It’s something that I’m not really looking forward to eventually. It’s like `Oh, right. I have to do that.' I’ve kinda lost interest in it.}

FP2 emphasized that the frequency of the messages also contributed to their disengagement. Specifically, receiving messages at sporadic intervals throughout the day could become overwhelming rather than helpful. Elaborating on the experience of interacting with several messages, FP11 stated:

\italquote{I just feel like if I don’t accomplish at least reading it and responding back, I’ve failed at something. \ldots So, it can get a little bit overwhelming.}

To enhance user engagement, participants provided suggestions for additional features. For example, they recommended implementing features for journaling or documenting daily events. They also suggested a more organized messaging interface to separate and retrieve distinct dialogues for future reference.

\subsection{Broader Social Contexts Underlying Disengagement} 

Unlike instances where engagement decreased gradually due to perceived shortcomings of the tool, there were often times when participants abruptly disengaged due to life circumstances that demanded their full attention and energy. In these instances, disengagement often happened despite satisfaction with the tool. We refer to these circumstances as situational engagement disruptors (SEDs), \topic{which are summarized in Table \ref{tab:sed_summary}}. We describe them in detail below.

\begin{table}[h!]
\centering
\caption{\topic{Summary of SEDs identified from Study 1}}
\label{tab:sed_summary}
\begin{renv}
    
\begin{tabular}{|p{4cm}|p{9cm}|}
\hline
\textbf{Category}                     & \textbf{Examples of Disruptors}                                                                                     \\ \hline
Academic and Workplace Responsibilities & Academic coursework,  phone restrictions, deadlines, and unpredictable schedule                         \\ \hline
Family Responsibilities                & Childcare, household tasks, and financial concerns                                   \\ \hline
Other Disruptors                      & Illness, relocation, sleep disturbances, and grief                                       \\ \hline
\end{tabular}
\end{renv}

\end{table}

\subsubsection{Academic and Workplace Responsibilities}

Several participants shared their experiences dealing with academic stress and its impact on their ability to engage with text messages. For instance, FP16 expressed feeling overwhelmed by their coursework, stating that the initial enthusiasm they had when signing up for the program was overshadowed by mounting academic pressures. As the semester progressed, FP7 reflected on the increasing number of assignments and exams that demanded their full attention. Both FP7 and FP16 noted that they derived value from the program by gaining insights into new psychological theories and considering how to integrate these learnings into their daily lives. However, the demands of their course load and the stress associated with final exams subsequently made it challenging for them to continue dedicating time for such reflection.

Participants also extended their list of challenges preventing them from engaging with text messages to workplace rules and responsibilities. Reflecting on their work environment, FP8 shared the following:

\italquote{Unfortunately, I work in an environment where you're not allowed your phone. \ldots It's like a secure room, so any text I get during the workday tends to all come together. I read them, but it's hard to engage with all of them, like, all at once.}

\noindent
Several participants highlighted categories of occupations that may inherently conflict with the ability to respond to text messages promptly. FP14 pointed out truck drivers as an example, explaining that the nature of their job may often prevent them from safely and conveniently engaging with text messages while driving. The irregular and unpredictable timing of their breaks compounds this issue as it complicates anticipating when they will have the chance to interact with such messages.

Even individuals who do not face strict rules or difficulties regarding phone usage found it challenging to prioritize self-care during working hours, stemming from the perception that taking a moment for personal wellbeing might detract from their focus on work. For example, FP2 noted that when their phone buzzes repeatedly with notifications, they feel compelled to redirect their attention from work to the messages. Although they found the messages useful for learning how to cope with stress, they were concerned about continually switching back and forth between work and text messages. They explained: 
\italquote{That experience of, you’re focused at work and you hear your phone go off -- it sounds like you have a reaction like ``Okay, what is it?'' You have to completely redirect your attention.}



The tensions between work tasks and personal wellbeing activities were highlighted further in FP4's experience. They indicated that they only felt compelled to promptly respond to messages that concerned urgent job responsibilities or impending work-related deadlines. Messages suggesting strategies to handle negative emotions did not hold the same sense of urgency or offer benefits that could surpass the importance of completing time-bound professional tasks.
Consequently, these messages were often relegated to a lower priority and could be easily forgotten or overlooked. 


\subsubsection{Family Responsibilities}
A few participants expressed feelings of overwhelm regarding their family responsibilities, most notably women who felt responsible for managing household duties. FP2 and FP6 shared their experiences as young parents and the impact it has on their ability to engage with text messages. FP2 explained:

\italquote{I think that [the text messaging program] was really helpful for me. \ldots But I'm in school, I'm working, my oldest is in kindergarten, my youngest is in daycare, so I'm all over the place. And part of what I struggle with is being – I'm also 25. \ldots When I have that free time, I never use it as `me time'. It's clean up the kids' room, get my laundry done, make sure everything's good so that when they get back home, we've got a new space to tear apart again. \ldots And it's never me just taking a nap. \ldots If I'm home alone, I'm getting homework done. I'm getting chores around the house done. \ldots It goes along with being a parent. All of your time is dedicated to other people's needs and not so much your own.}

\noindent
Caring for children and other family members had a significant impact on people's ability to plan and set personal goals. The demands of managing household duties, childcare, and other familial obligations can create a sense of overwhelm and restrict people's capacity to allocate time and energy for reflection or self-care. Hence, FP6 emphasized that young parents would benefit from adding more structure to their lives, whether it be through creating routines or setting aside dedicated time for self-care activities. 


For some, anxiety and overwhelm centered on their families' financial stability. FP17 described ongoing worries and efforts around maintaining a balanced budget while meeting their family's basic needs, noting that such worries could easily lead people to deprioritize engaging with text messages. They commented:

\italquote{Even when I was busy, I would take the time \ldots People are busy, and a lot of people are just not gonna take the time to do it depending on how depressed the person is. They've got bills. They've got other things. The way things are right now, people are struggling just to make it. So, it’s gonna be a challenge. People are thinking, 'I’ve got a family to feed and if I don’t go do this, this, and this, I’m not gonna \ldots surmount any of the challenges'.}

\noindent
This preoccupation with financial challenges consumed their thoughts to such an extent that it became difficult for them to redirect their focus to other areas of life, namely self-care. Despite recognizing the potential benefits of the program's messages, participants sometimes found themselves unable to fully engage with them.


\subsubsection{Other Disruptors}

Outside of academics, work, and family, participants highlighted other circumstances that disrupted their usual routines and hindered their engagement with text messages. These disruptions encompassed a range of unexpected events and situations, including brief illnesses, changes in residence, and the death of loved ones. Throughout the eight-week period, some participants reported experiencing health issues like COVID-19 and bronchitis that impacted their mood and energy levels, making it difficult to engage with the messages.
FP8 revealed that missing messages for a few days during their COVID-19 illness led them to disengage from the program altogether as the perceived difficulty in catching up with the missed messages seemed daunting. They stated:

\italquote{Getting a text is just actually really nice. \ldots When I got COVID and I was getting messages \ldots that kind of thing skews my engagement with it a little bit because I was too tired to even read my phone let alone actually engage with anything. And sometimes, it is annoying when you get several messages all at once.}

Meanwhile, participants like FP5 and FP8 shared how insomnia and changes in their sleep schedule disrupted their ability to interact with the messaging program. They relayed that sleep deprivation led them to a state of heightened busyness yet diminished presence in their daily activities. This lack of energy and focus negatively impacted their engagement with text messages, making comprehension and response more challenging. FP5 echoed a similar sentiment by underscoring the importance of establishing a regular sleep routine to better manage their interaction with the program.


Additionally, participants highlighted the difficulties they faced when adapting to uncertain and unfamiliar housing environments due to new job opportunities. FP12 mentioned that they had to constantly move between two cities for work. The uncertainty surrounding their living arrangements consumed their thoughts and hindered their ability to engage with text messages. Participants also mentioned personal, unanticipated reasons for disengagement. FP4, for example, shared the experience of losing a close friend on the same day they enrolled in the program. The grieving process consumed their emotional energy, leading them to ignore the messages for the first few days.

\section{Procedure of Study 2}

The first study identified several critical SEDs within the context of engagement with a DMH tool. These SEDs emerged as broader factors originating from the social contexts surrounding an individual. Spurred by these insights, our subsequent efforts were directed towards identifying potential solutions to address and alleviate these SEDs. To this end, our second study involved conducting design workshops with \topic{25 participants} aimed at generating ideas for further exploration. \topic{While we drew inspiration from the deployment of a text messaging tool for these workshops, we aimed to explore how any sort of DMH tool can be designed to address SEDs.}
This phase of the study was approved by \redacted{the University of Toronto's} Research Ethics Board, and its logistics are described below. 

\subsection{Participants}

We utilized promotional calls across various social media platforms to recruit diverse participants who felt overwhelmed by their life circumstances. 
The study was framed as a collaborative design activity to understand challenges in everyday life that could impede engagement with DMH tools and to explore strategies for addressing these challenges. 

\topic{In the first study, participants with elevated symptoms of depression and anxiety frequently highlighted feeling overwhelmed by competing demands \cite{kabigting2019conceptual, hopps1995power} as a primary barrier to sustained interaction with DMH tools. Recognizing that this challenge is also prevalent in populations without a formal diagnosis or clinically elevated symptoms, we expanded the focus in the second study to include other populations that would benefit from DMH tools. This second study included individuals experiencing elevated stress irrespective of their current levels of depression or anxiety symptoms, allowing us to incorporate their diverse perspectives in our workshops.}

Our eligibility criteria required participants to self-report moderate to severe levels of stress according to the Perceived Stress Scale (PSS) (determined by a score of 14 or more) \cite{cohen1994perceived}.  Eligibility criteria also required participants to be at least 18 years old and residents of North America. 
Interested individuals could follow a link to learn more about the study online. After completing a brief eligibility survey, eligible participants were asked to provide informed consent.

\topic{The final cohort for this study consisted of 25 participants} with an average age of $25.0\pm4.2$ years old. Participants identified with multiple genders (15 men, 9 women, 1 non-binary) and multiple racial groups (9 Asian, 7 White, 6 African American, 1 Native Hawaiian/Other Pacific Islander, and 2 mixed race). 
\topic{Similar to the first study, participants were asked whether they had ever been diagnosed with major depression or anxiety disorder by a specialist or healthcare provider, although a formal diagnosis of depression or anxiety was not required for participation.
Nine participants (36.0\%) reported being diagnosed with depression, and 12 participants (48.0\%) reported being diagnosed with an anxiety disorder.} We refer to these participants as SP1--SP25. 

\subsection{Data Collection and Procedures}

We hosted five online design workshops through the Google Meet videoconferencing platform, with each accommodating 4–6 participants and two research team members. Each participant attended only one session, and every workshop lasted approximately 90 minutes. Participants were compensated \$30 USD for their participation.

The workshops began with a brief introduction to the workshop's goals, followed by a presentation from a research team member detailing the findings of our first study. After this overview, participants shared their personal experiences with DMH tools similar to the Small Steps SMS program and their general thoughts about using technology (e.g., apps, text messaging) to support mental health self-management. \topic{They were also asked to reflect on SEDs they had experienced or considered relevant to their interactions with DMH tools, both within and beyond the findings of the first study.} We then presented them with a series of hypothetical scenarios distilled from common issues that emerged from the formative study. Each scenario presented a set of circumstances in which an individual might become overwhelmed, spurring participants to reflect on the challenges one might face while trying to engage with a DMH tool as intended. \topic{These scenarios are summarized in Table~\ref{tab:scenario} in Appendix \ref{app: scenarios}.}

Following the hypothetical scenarios, participants engaged in three rounds of activities during which they documented their thoughts in online documents and shared their insights verbally. Each of these rounds lasted 15--20 minutes and revolved around one of the following questions:

\begin{enumerate}
\item Reflecting on times when you faced similar challenges, which self-care practices did you find most beneficial? What specific strategies would you recommend for managing such challenges in the future?
\item In what ways do you believe programs like the Small Steps SMS Program can be enhanced to help people adopt these self-care practices, especially during challenging times?
\item Are there strategies that you feel are broadly applicable regardless of specific life challenges? 
\end{enumerate}

\topic{We clarified that participants did not need to limit their recommendations to modifications for the Small Steps SMS program. They were encouraged to share insights that could be broadly applicable to the design of various DMH tools.}

\subsection{Data Analysis}
After consolidating the workshop transcriptions and participant notes in the online documents, we employed the same thematic analysis procedures that we utilized in the first study, although a distinct codebook was developed for this phase.

\subsection{Ethical Considerations}

We used similar measures to the ones detailed in Section \ref{study1_ethics} in order to ensure that participants were treated in an ethical manner. All researchers who spoke with participants were prepared to conduct the Columbia-Suicide Risk Assessment protocol, yet no risks emerged and no follow-ups were necessary.
\section{Findings from Study 2}

\topic{Similar to the first study, participants shared varied interpretations of DMH tools.
Participants mentioned enrolling in text messaging services (DP1, DP7, DP13, DP14), participating in digital journaling (DP1), engaging in social media channels (DP4, DP8, DP20), connecting with therapists over digital platforms (DP12), and using apps like Headspace (DP21), Calm (DP13), and Forest (DP5).}
They recommended several ways to account for the SEDs in the design of DMH tools. We describe these potential solutions below.

\subsection{Prioritization of Responsibilities}
During the workshop, participants identified that feelings of overwhelm often arise from juggling numerous personal and professional responsibilities. For instance, SP11 recalled how they had to balance multiple exams, complete their thesis, and work at a part-time job to support their household during their final year as an undergraduate student. 
Reflecting on these diverse experiences, participants acknowledged the crucial need to prioritize responsibilities. SP6 suggested that DMH tools could help individuals with multiple demands on their time as follows:

\italquote{Try to weigh all of these responsibilities and see which of them take precedence over the rest, and try to fulfill those responsibilities in that order. Prompt the user to deeply think about these responsibilities so they can get a better handle on things.}

Tied into these discussions was a recurring theme of procrastination. Participants often cited an accumulation of tasks that were deferred out of anxiety or a lack of motivation.
To mitigate this issue, participants like SP7 proposed that DMH tools could help users deconstruct larger responsibilities into smaller, more manageable tasks. They believed that this approach would make tasks more actionable, foster a sense of productivity, and subsequently disrupt the cycle of procrastination.
However, participants in the workshop recognized that prioritizing responsibilities often necessitates carefully considering and endorsing a rationale for why certain tasks take priority. They thought this could be facilitated by soliciting users' short-term and long-term objectives, and aiding them in formulating action plans to meet these goals. 
SP1 noted that seeing their responsibilities in written form \textit{`just makes it feel like grounded and like more in control'}.

Participants were also aware of the mental effort and cognitive resources required for these practices, making them more suitable for periods of relative downtime like weekends and holidays. SP7 suggested that DMH tools could help users articulate goals and plan actions during these periods. To keep these goals at the forefront, participants recommended reminders throughout the week, with SP7 emphasizing morning reminders to align daily activities with long-term objectives.

Still, SP21 cautioned that writing down detailed, elaborate plans on mobile devices could contribute to additional cognitive overload, even during free time. Instead, they suggested that users could be offered the option to either formulate their plans mentally or using traditional pen-and-paper methods.

\subsection{Incorporating Self-Care into Daily Routines}

Participants emphasized the importance of maintaining a balanced lifestyle with respect to sleep, diet, and exercise. This was regarded as an important factor in managing one's overall wellbeing, especially during periods of stress or overwhelm. Based on challenges identified in our first study related to maintaining a consistent sleep schedule in the face of SEDs, workshop participants proposed that DMH tools should include features to encourage sleep hygiene. SP10 suggested that users' preferred sleeping hours should be gathered during the registration process, allowing DMH tools to send prompts when individuals' activities on their phones continue beyond the specified bedtime.
During periods of illness or heightened stress, participants acknowledged that neglecting meals had an adverse effect on their mental health. In such scenarios, the recommendation was to focus on ensuring users remember to eat and hydrate. SP7 commented:

\italquote{Instead of like constant messages, add more general reminders like eating, drinking water, etc. Depending on the time of the day, you can send messages like `have a snack,' `drink some water,' `do breathing exercises,' or food and snack ideas.}

Reflecting on the toll of substantial familial responsibilities, participants like SP23 and SP25 expressed a tendency to neglect self-care in favor of attending to the needs of others. They often attributed this inclination to the absence of a systematic approach to time management. To address this issue, they suggested that DMH tools could assist users in carving out dedicated time for self-care. SP23 elaborated that these reserved times could be spent meeting friends or engaging in restful activities like reading or sleeping. However, we observed varied preferences for how this personal time should be distributed; while some preferred an hour each day, others leaned towards larger blocks of time or even full days every few weeks.

To support users in navigating their daily activities and appropriately allocating time for self-care, participants like SP5 and SP14 suggested that DMH tools should consider integrating with users' digital calendars or other platforms commonly used for academic or professional communication (e.g., Microsoft Teams, Slack). They proposed that this synchronization would allow such tools to gain a nuanced understanding of users' schedules and obligations, consequently enabling more personalized and timely interventions. For example, reminders or prompts for self-care activities could be strategically scheduled during free periods identified in users' calendars, thus assisting them in maintaining a healthy work-life balance amidst their professional or academic responsibilities. Participants also anticipated that seeing a time slot reserved for self-care in their calendar would reinforce their commitment to these activities.

\subsection{Alternate Framing of Disengagement}

Participants acknowledged that irrespective of the quality and effectiveness of a DMH tool, users are likely to experience occasional disengagement due to either voluntary or involuntary reasons. They highlighted the mental hurdle of re-engaging with these tools after a period of non-use. Reflecting on their past experiences with other DMH tools, SP1 expressed a negative sentiment regarding the tools' emphasis on continuous use. They noted that when they took a break from the tools, the feeling of losing their streak demotivated them. This demotivation stemmed from the perception that they would have to put in significant effort to regain the level of consistent use they had previously achieved. They commented:

\italquote{If I consistently used it every day and then I forgot one day, or like I forgot a couple of times, I'd be like, `Oh, but I already lost the streak.` Then I might as well just like give up the whole thing, which is not great.}

\noindent
To support users during periods of heightened stress or activity, SP2 suggested enabling them to specify days when they might be occupied with exams or work responsibilities. Marking such days in a DMH tool would prompt the system to lower engagement expectations, helping prevent users from completely withdrawing. For example, users could be asked to reflect on their mood or energy level with a number or emoji rather than a full journal entry. 

Despite these recommendations, participants acknowledged that occasional disengagement is inevitable.  
During prolonged periods of user inactivity, SP22 proposed that DMH tools should respond by issuing straightforward and actionable suggestions to re-engage the users. However, SP5 and SP9 warned that the frequency of such suggestions should not be overwhelming, as too many notifications could potentially irritate users and cause them to develop a strong aversion to the tool. 
SP9 shed light on this perspective from their previous experiences:

\italquote{From the meditation app, I actually got a little bit annoyed because it would sort of bloat my notification list. I kind of get mad if I don't see the value in the amount of times my phone is being blown up. \ldots If it's sending me a lot of messages and long messages, I might go into rebellious mode and just not even answer out of pure hate or maybe just possibly hit skip.}


In light of these reflections, workshop participants suggested a shift in perspective. Instead of perceiving periods of disengagement as user failures, DMH tools should understand and accept these moments as part of the user's journey. SP6 highlighted the importance of acknowledging struggles and promoting the idea that focusing on more controllable aspects of life, such as engagement with a DMH tool, can positively influence one's mental health. 
The goal here would be not to pressure users into continuous usage, but to foster an environment where they feel supported and understood, even during periods of disengagement. Participants emphasized that the success of using DMH tools should not be solely defined by continuous usage but rather the fact that users are choosing to utilize the tool to improve their wellbeing, despite barriers and setbacks.

\subsection{Leveraging External Support and Community Resources}
Participants underlined the significant role external support and community resources can play when it comes to managing overwhelming situations. 
Drawing on their academic experiences, participants like SP2 and SP8 expressed the ease with which people can overlook potential help from their peer network. They described times when they grappled with academic assignments on their own, forgetting they could lean on friends and teaching assistants for guidance. They suggested that a gentle reminder to consider support resources could be beneficial in moments of high stress. Participants also spoke about how they appreciated reading narratives that echoed their specific struggles. They expressed that seeing examples of how others have managed to navigate similar issues could offer them hope, strengthen their belief in their ability to improve their situation, and potentially provide practical directions to explore.

On the whole, participants described that addressing individuals' psychological wellbeing sometimes must go beyond merely providing supportive messages or teaching psychological strategies. In addition to encouragement to draw on their existing social networks for support, some participants suggested that people dealing with financial burdens may benefit from DMH tools that direct users to unconventional financial resources. Drawing upon a hypothetical scenario of a single mother juggling multiple jobs and having little time for self-care, SP13 posited one such example:

\italquote{For single mothers, there are many government and non-government organizations that give grants. \dots And if possible, she could collect grants, or maybe someone wants to start up a small business.}

\noindent
Concerning individuals with unstable housing conditions, participants pointed to websites and social media groups dedicated to helping people find suitable housing. They relayed that many users are often unaware of these resources, so DMH tools could provide a convenient gateway to them. 

Participants recognized the difficulty in tailoring DMH tools to users' specific ongoing life challenges, as these issues are often open-ended and unique to each individual. Therefore, 
SP10 proposed that DMH tools could learn from the approach employed by the social media platform Snapchat, which collects data on users' preferences and interests as part of its sign-up process. They went on to say: 
\revision{\italquote{When you first log in as a user, you get the option to choose what kind of problem you're struggling with, kind of the scenario you're in. And you know, you can get responses based on that scenario.}}

In a complementary suggestion, SP1 recommended that DMH tools could then compile a repository of resources tailored to address those challenges. When users express their concerns, a matchmaking process could be initiated to direct people to resources related to their issues. Given recent excitement for large language models (LLMs), a few participants theorized that LLMs could be used to match people's open-ended concerns to resources. Nevertheless, participants were wary of the reliability, suitability, and scalability of such a process.
\section{Discussion}

Our work offers evidence and actionable insights for the HCI and CSCW communities, emphasizing the essential need to consider the broader social contexts in which DMH tools operate. \topic{Although our observations originate from a text messaging tool used for self-managing depression and anxiety symptoms, the SEDs we identified likely have similar impact on the usage of other DMH tools. For instance, several prior studies have reported that social circumstances can pose barriers to people's engagement with DMH mobile apps and online programs \cite{borghouts2021barriers, bhattacharjee2024exploring, harjumaa2015user, bhattacharjee2023investigating}.} We enrich the existing research by exploring how social contexts affect interactions with DMH tools, highlighting the inherently sporadic nature of engagement. Our proposed solutions—community support, socio-technical interventions, and individual goal-setting—aim to redefine successful engagement by adapting to varying user involvement levels and addressing specific SEDs.


In the following discussion, we first outline how our findings respond to our research questions and contribute to existing literature on engagement with DMH tools. We then discuss the limitations of our study.

\subsection{RQ1: Influence of SEDs in Dictating Engagement with DMH Tools}
Prior literature on disengagement has emphasized individual correlates like motivation or symptom severity \cite{lipschitz2023engagement, doherty2018engagement, venkatesh2003user, marangunic2015technology, miller2016game, yoo2024missed} and the ways in which disengagement may reflect misalignment between a tool and one's preferences \cite{bhattacharjee2023investigating, jardine2023between, perski2017conceptualising}. 
\topic{Previous researchers have also commented on how circumstantial constraints like job-related issues and financial hardships frequently lead to termination of traditional in-person therapy \cite{renk2002reasons, roe2006clients, hynan1990client, carolan2018employees}.
Our work extends this literature by identifying contextual constraints from SEDs that instigate disengagement with DMH tools. We observed that these constraints significantly influence both subjective (e.g., perceived commitment and burden) and behavioral (e.g., interaction frequency, long-term adherence) measures of engagement. While the SEDs identified in our study emerged from one specific intervention, they reflect broader engagement challenges that transcend any given technology. These issues can arise in interactions with different types of DMH tools including mindfulness apps \cite{laurie2016making}, digital self-care platforms \cite{borghouts2021barriers}, or tools designed to manage symptoms of depression and anxiety \cite{jardine2023between, bhattacharjee2023investigating}}. We observed that SEDs can make it challenging for users to maintain consistent engagement with DMH tools even when they command as little effort as reading a text message. 

\subsubsection{The Role of Feeling Overwhelmed in Disengagement}
Our investigation uncovered a handful of diverse SEDs that can influence engagement with DMH tools: academic challenges, workplace expectations, familial duties, and unanticipated life events. Although DMH tools are often theorized to overcome circumstantial constraints because of their convenience and accessibility, we demonstrate that SEDs remain significant impediments to sustained engagement. \topic{Echoing previous research \cite{borghouts2021barriers, carolan2018employees, bhattacharjee2023investigating}, we observed that consistent engagement faltered when individuals found it difficult to weave these digital interventions into their everyday routines.} When demands exceeded coping capacity, individuals often felt overwhelmed \cite{kabigting2019conceptual, hopps1995power}, leading to the deprioritization of self-care activities like DMH tools in favor of immediate or tangible tasks.


\topic{Further complicating this situation were unexpected life events and disruptions, ranging from health issues and fluctuations in sleep patterns to relocations, all of which have been found to impact engagement with DMH tools \cite{borghouts2021barriers, karlgren2023sleep, lattie2020designing, perski2017conceptualising}. This cascade of challenges hindered people's ability to plan and set priorities, contributing to heightened distress.} Consequently, self-care activities could be perceived as additional burdens, even when participants acknowledged their value and expressed a genuine affinity for them.


\subsubsection{Handling Disengagement}
Once participants experienced voluntary or involuntary disengagement with a DMH tool, they frequently harbored the belief that they had irrevocably fallen behind. \topic{As prior work on mental health apps has also acknowledged \cite{six2021examining}, the notion of `falling behind' or `losing a streak' might evolve from a simple logistical concern into a formidable emotional barrier.} This perception weakened their resolve to continue with self-care efforts, initiating a cycle where re-engagement with the tool became increasingly difficult. The pressure to maintain a streak occasionally transformed into a reminder of their shortcomings, leading to self-disappointment \cite{bekk2022all, toda2017dark, hadi2022gamification}. \topic{These observations contribute to technology adoption models \cite{venkatesh2003user, dwivedi2019re} by underscoring how emotional barriers can shape users' motivation and perceived usefulness of DMH tools.} 


The revelation that these broader challenges often underlie individuals' disengagement suggests that a more nuanced and comprehensive understanding of users' lives is imperative for crafting engagement strategies that genuinely support psychological wellbeing \cite{pendse2022treatment, allen2014social, slavich2020social, murnane2018personal, kaziunas2017caring}. \topic{Since engagement can often be sporadic, DMH tools should accommodate varying levels and types of interaction without penalizing users or creating additional stress. Taking inspiration from slow and temporal design philosophies \cite{grosse2013slow, pschetz2018temporal}, DMH tools should prioritize thoughtful, user-paced interactions to help users achieve long-term goals over time. They could include flexible goal-setting activities to integrate self-care into users’ daily lives or targeted support for managing the underlying causes of overwhelm; we elaborate upon some potential design solutions in Section~\ref{subsec: RQ2}. DMH tools could also offer customizable schedules that adapt to users’ unique routines, especially during periods of overwhelm. At the same time, it is important to avoid introducing unpredictability that could exacerbate overwhelm, as expressed by participants like FP11. DMH tools should also ensure that flexibility and adaptability are transparently communicated and predictable in their operation, so users feel supported rather than disrupted.} 

\subsection{RQ2: Designing and Implementing DMH Tools to Account for the Social Contexts}
\label{subsec: RQ2}

We recognize the complexity involved in proposing design solutions that respond to diverse SEDs. Even though participants across both studies came from varied life circumstances, they shared common experiences of overwhelm and disengagement and largely found similar types of solutions to be valuable. \topic{As summarized in Table \ref{tab:socio_technical_solutions}, these suggestions were often socio-technical rather than purely technological, extending frameworks like Social Safety Theory \cite{slavich2020social} by demonstrating how DMH tools can alleviate perceived social threats. In doing so, they open up new discussions on how DMH tools can foster a sense of social safety within users' daily interactions. We elaborate on these ideas below.}

\begin{table}[ht]

\caption{\topic{Socio-technical solutions based on our findings}}
\label{tab:socio_technical_solutions}
\centering
\begin{renv}

\begin{tabular}{|p{3cm}|p{3cm}|p{7cm}|}
\hline
\textbf{Theme} & \textbf{Action} & \textbf{Description} \\ \hline
\multirow[t]{3}{3cm}{Prioritizing Self-Care Through Structured Goal-Setting} 
& \raggedright Balancing Self-Care and Responsibilities & Encourage users to prioritize self-care alongside personal and professional responsibilities by offering structured prompts during quieter periods, such as weekends or evenings, to help reduce perceived overwhelm. \\ \cline{2-3}
& \raggedright Decomposition of Tasks & Simplify complex goals by breaking them into smaller, actionable steps, which makes achieving tasks more manageable and less daunting for users. \\ \cline{2-3}
& \raggedright Facilitating Long-Term Planning & Encourage the development of plans that align with users' broader life goals, supported by interactive features that aid in strategic planning. \\ \cline{1-3}

\multirow[t]{3}{3cm}{Designing for Flexible Engagement} 
& \raggedright Adaptive Engagement Strategies & Design tools to accommodate varying levels of user engagement, offering simple, low-effort options during high-stress periods and more intensive, high-effort activities during times of greater user availability. \\ \cline{2-3}
& \raggedright Establishing Foundational Support & Foster a consistent relationship between users and DMH tools by providing unintrusive reminders during low-engagement periods and allowing flexible, pressure-free re-engagement opportunities. \\ \cline{2-3}
& \raggedright Reframing Disengagement & Normalize occasional disengagement by emphasizing the role of DMH tools as a natural part of the user journey, thereby reducing feelings of guilt or failure. \\ \cline{1-3}

\multirow[t]{4}{3cm}{Addressing SEDs Within Social Contexts} 
& \raggedright Targeted Socio-Economic Support & Integrate pathways to practical resources, including financial counseling, housing assistance, and grants for vulnerable populations. \\ \cline{2-3}
& \raggedright Peer and Community Integration & Enable tools to foster a sense of belonging by sharing relatable narratives and connecting users with peer or community-based support systems. \\ \cline{2-3}
& \raggedright Multi-Stakeholder Collaboration & Collaborate with financial technology companies, local banks, and other institutions to provide accessible workshops or tools addressing life complexities. \\ \hline
\end{tabular}
\end{renv}

\end{table}


\subsubsection{Prioritizing Self-Care Through Structured Goal-Setting}

In line with previous studies \cite{bhattacharjee2022kind, brown2014health, chaudhry2022formative, jardine2023between, epstein2015lived}, our findings underscore the difficulty users encounter when trying to weave self-care activities seamlessly into their daily lives. Conflicts arise when people's presumed responsibilities interfere with the timing and perceived importance of these activities, calling for tools that assist people in prioritizing their myriad responsibilities. Our findings indicate that structured goal setting, often a standard objective of many therapeutic interventions \cite{chand2023cognitive, agapie2022longitudinal, epstein2015lived}, holds significant importance in general wellbeing management when engaging with DMH tools. It becomes imperative for these tools to consider the broader goals of the users, thereby ensuring that the activities proposed are in harmony with, and not contradictory to, those goals. 

\topic{These findings highlight the critical role of DMH tools could play in counteracting the tendency people have to sideline self-care during busy periods.}
To prevent academic or professional obligations from continually overshadowing self-care activities, DMH tools could assist users in setting aside designated times for self-care while emphasizing how self-care activities directly contribute to personal objectives. These personalized goals could correspond with their broader life objectives, highlighting the real-world implications of self-care practices \cite{jennings2018personalized, lindhiem2016meta, agapie2022longitudinal, epstein2015lived}. Incorporating visual analytics tools as suggested in health and wellbeing visualization studies \cite{baumer2014reviewing, kocielnik2018reflection} could provide users with graphical representations of improvements to their wellbeing, linking them to enhanced performance in other life areas. Further reinforcement can come from integrating educational modules that elucidate the scientific connections between wellbeing practices, work performance, and overall life improvement \cite{howe2022design, bhattacharjee2023understanding}. 

In line with literature on collaborative goal-setting \cite{agapie2022longitudinal, lee2021sticky, zakaria2019stressmon, jung2023enjoy, xu2023technology}, DMH tools can also serve as a collaborative partner in sorting through overwhelming obligations. For example, DMH tools could lead users through guided conversations that help them decompose broad objectives into distinct, actionable steps \cite{o2018suddenly, bowman2022pervasive, borghouts2021barriers}. Tools could also help users identify which responsibilities are essential, which could be considered deferred, and which could be ignored altogether. Such dialogues would help users allocate their time wisely and concentrate on high-priority activities \cite{pereira2021struggling, grover2020design}.
Recognizing that these guided interactions might require dedicated time and thoughtful reflection, participants in our study suggested that users should be encouraged to complete these brief exercises during their idle time \cite{poole2013hci, bhattacharjee2022design}. Idle time could include moments when individuals are engaged in low-attention activities, such as checking messages or browsing social networking platforms. Once a plan has been crafted, DMH tools could then offer occasional reminders and reflection prompts, helping users stay on track with the incremental steps needed to achieve their overall goals. \topic{This form of technological mediation would allow self-care practices to coexist with other responsibilities, ensuring that these essential activities receive the attention they deserve.}

\topic{However, the potential for collaboration between users and DMH tools can be disrupted by a multitude of factors \cite{flathmann2024empirically, caldwell2022agile}. Power imbalances may emerge when the tool imposes too much control over the user’s decision-making process, potentially diminishing their sense of personal agency \cite{flathmann2024empirically, bhattacharjee2023informing}. Furthermore, trust issues can severely disrupt collaboration, whether they stem from a sudden recommendation that undermines the tool's reliability or from concerns about the security of personal data \cite{caldwell2022agile}. Future studies should explore the landscape of human-DMH tool collaboration to identify these disruptors as well as strategies to mitigate them.}





\subsubsection{Designing for Flexible Engagement}

Participants in our study expressed a strong desire for a reframing of what constitutes successful engagement with DMH tools. Although the core intent of DMH tools is to foster wellbeing, an emphasis on sustained engagement can inadvertently defeat this purpose \revision{even when such engagement is necessary.} Our research underscores that a relentless push for streaks and frequent notifications can introduce unnecessary stress, echoing commentaries against over-gamification \cite{etkin2016hidden, bekk2022all, liu2017toward, jia2016personality, hamari2014does, epstein2015lived}. 
To address this, DMH tools can consider offering options for taking short breaks without jeopardizing accumulated progress. Additionally, these tools should normalize brief periods of disengagement and formulate strategies to encourage users to return and continue their wellbeing journey.

Nevertheless, disruptions are inevitable, so it is equally important to identify appropriate cadences for re-engaging users. Our findings suggested that participants are divided when considering how often a system should reach out to a disengaged user, as efforts to re-engage users could risk further demoralizing or annoying them \cite{bhattacharjee2022kind, epstein2015lived}. A promising strategy could be to frame acts of re-engagement after a hiatus as a meaningful achievement. Rather than dwelling on periods of inactivity, DMH tools should thus focus on fostering a foundational relationship with users, utilizing unintrusive reminders during periods of low engagement as a way to encourage continued interaction~\cite{bhattacharjee2023integrating, yardley2015person}. \topic{This foundational relationship can play a critical role in helping users maintain a consistent practice of self-care. When this relationship is absent, individuals may be forced to entirely rely on their own discipline and time management abilities;  if overwhelmed, the lack of external support can lead to sporadic or neglected self-care. Therefore, it is vital for DMH tools to regularly foster this relationship, allowing users to flexibly engage without undue pressure.}

\topic{We acknowledge the challenges in maintaining this relationship as integrating personal, academic, or work schedules with supportive resources in DMH tools may shift them from lightweight, low-burden systems to potentially more complex, high-burden ones.} 
\topic{Approaches like causal pathway diagramming \cite{klasnja2024getting} can help identify mechanisms to achieve a balance between these extremes. DMH tools could dynamically adjust their functionality based on user context, factoring in preconditions such as user availability and moderators like their energy levels. During high-stress periods or busy workweeks, the system could offer lightweight support that does not require a response or involves only quick and simple tasks \cite{bhattacharjee2023investigating}. These low-effort interactions can aim to provide immediate support, potentially facilitating proximal outcomes like stress relief or a temporary mood boost. Conversely, when users have more energy or availability (e.g., idle time \cite{poole2013hci, bhattacharjee2022design}), the tool should encourage intensive activities like goal-setting or reflection to achieve distal outcomes such as overcoming life stressors or reaching professional goals \cite{kornfield2020energy}. Such an adaptive model should ensure all interactions, regardless of effort level, align with users' broader life objectives.}

\subsubsection{Addressing SEDs Within Social Contexts}

Participants also pointed to the need for DMH tools that help them directly manage the specific SEDs underlying their overwhelm and disengagement. \topic{They described that tools should assess the specific SEDs individuals were facing and, where possible, match them to solutions \cite{ma2023contextbot, jardine2023between, zakaria2019stressmon, fang2022matching}.}  For example, targeted digital interventions could address challenges in carrying out family obligations by providing access to virtual family counseling and online parental education to help individuals navigate the complexities of familial responsibilities \cite{chi2015systematic, smout2023enabling}. To better manage finances, digital financial literacy workshops and access to personal finance management tools might empower individuals to take control of their financial situations, mitigating anxiety and potential mental health deterioration \cite{boyd2016earned, mehra2018prayana}. Collaborative endeavors with financial technology companies or local banks could facilitate these online workshops, providing a practical approach to managing financial complexities \cite{dillahunt2022village, moulder2014hci}. Moreover, online community support groups could provide digital resources for childcare, eldercare, or other family-related needs in order to alleviate some of the burden on those balancing both family and work responsibilities \cite{gisore2012community, chaudhry2022formative, butler2012relationship, compton2015social, smout2023enabling}. Thus, past works emphasize the necessity for multi-level interventions that extend across and address various life dimensions and societal factors \cite{xu2023technology, sallis2015ecological, scholmerich2016translating, bauer2022community, murnane2018personal}. 


Previous research in implementation science underscores the importance of the processes through which technologies are disseminated and delivered to users for effective psychological wellbeing support \cite{graham2020implementation, hermes2019measuring, powell2012compilation, powell2015refined}. Whether in educational environments or workplaces, numerous opportunities exist for enhancing the integration of psychological wellbeing support within existing technological frameworks \cite{graham2020implementation, haldar2022collaboration, xu2023technology}. Interventions can be introduced at opportune moments using digital platforms such as email and calendars \cite{howe2022design}. 
Furthermore, \citet{graham2020implementation} recommend strategies like providing access to educational materials to help users learn to use a tool and designated contact persons for assistance. It should be noted that support does not necessarily have to emanate solely from mental health professionals; trained peers and paraprofessionals can periodically extend their assistance, thereby fostering a sense of accountability and support \cite{sultana2019parar, anvari2022behavioral, okolo2021cannot, burgess2019think, jardine2023between, haldar2022collaboration}.

\subsection{Limitations and Future Work}
Our work has a few limitations. First, our research focused on the experiences of individuals residing in North America. A more diverse participant pool from various geographical and cultural backgrounds might have revealed additional ones. 
\topic{For instance, norms around workplace and educational practices vary significantly across cultures \cite{southard2011vacation, parekh2022cross}. Some countries have regular short breaks embedded in their work culture while others do not, and such norms could influence people's willingness to interact with DMH tools during work hours. Even within regions of the same country, workplace practices and norms can differ \cite{parekh2022cross}. Future research should encompass diverse populations, including those both within and beyond North America, to explore how varying cultural and environmental contexts impact engagement with DMH interventions.}

Second, the specific context of our study — such as its duration and focus on a text messaging program — may limit the generalizability of our findings. Many of our insights should be applicable to the broader usage of DMH tools, yet different intervention designs might reveal distinct disruptors. \topic{Restricting the second study to participants with elevated depression and anxiety symptoms, as in the first study, might have provided findings more tailored to that population. Future studies could also explore how prior DMH tool usage influences future engagement with other tools.}

Third, we acknowledge that the optimal level of engagement with DMH tools is a complex topic that intertwines with the clinical model underlying the tool itself \cite{knights2023framework, saleem2021understanding, yardley2016understanding}. We also recommend that the proposed solutions from the second study should undergo thorough validation and be implemented in collaboration with clinicians and domain experts. Our research does not suggest that all DMH tools necessitate or should strive for continuous, long-term engagement. Rather, \topic{our insights are particularly relevant to DMH tools designed for extended engagement over weeks or months, as interventions requiring only brief user involvement, such as single-session interventions \cite{schleider2020future}, are less likely to be impacted by the disruptors we identified. Therefore, we limit our claims to tools intended for long-term use and do not extend them to short-term DMH interventions.}
\section{Conclusion}
\revision{Factors stemming from one's social context can play a critical role in shaping psychological wellbeing. Prior studies have noted that these factors may also impact the extent to which individuals engage with DMH tools. In our study, we explore how an individual's social context affects their engagement with a text messaging program for managing depression and anxiety symptoms.} Through interviewing participants at two time points during the eight-week program, we discovered that people experienced various SEDs ranging from personal obligations and professional duties to unexpected health concerns. Subsequent design workshops involving 25 participants generated practical solutions for addressing these SEDs. These include introducing structured approaches to goal-setting, reframing the concept of disengagement, and leveraging external resources. Our findings prompt new viewpoints into how engagement should be understood and managed, providing actionable recommendations for the design of future DMH tools.




\bibliographystyle{ACM-Reference-Format}
\bibliography{sample-base}
\newpage
\appendix

\section{Flow Diagram of Text Message Dialogues}
\label{sec: flow}

\begin{figure}[h]
    \centering
    \begin{subfigure}[b]{0.42\textwidth}
        \centering
        \includegraphics[width=\textwidth]{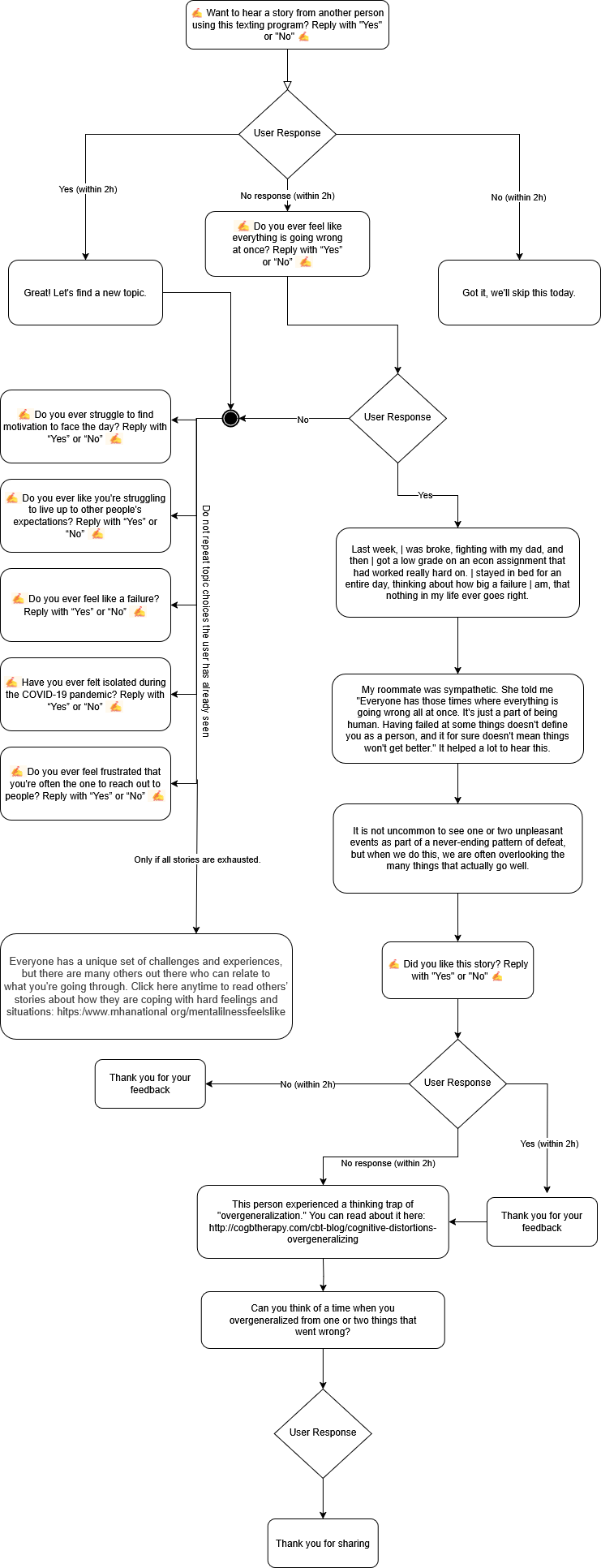}
        \caption{\topic{A message dialogue to communicate a relatable challenging experience}}
        \label{fig:pic1}
    \end{subfigure}
    \hfill
    \begin{subfigure}[b]{0.53\textwidth}
        \centering
        \includegraphics[width=\textwidth]{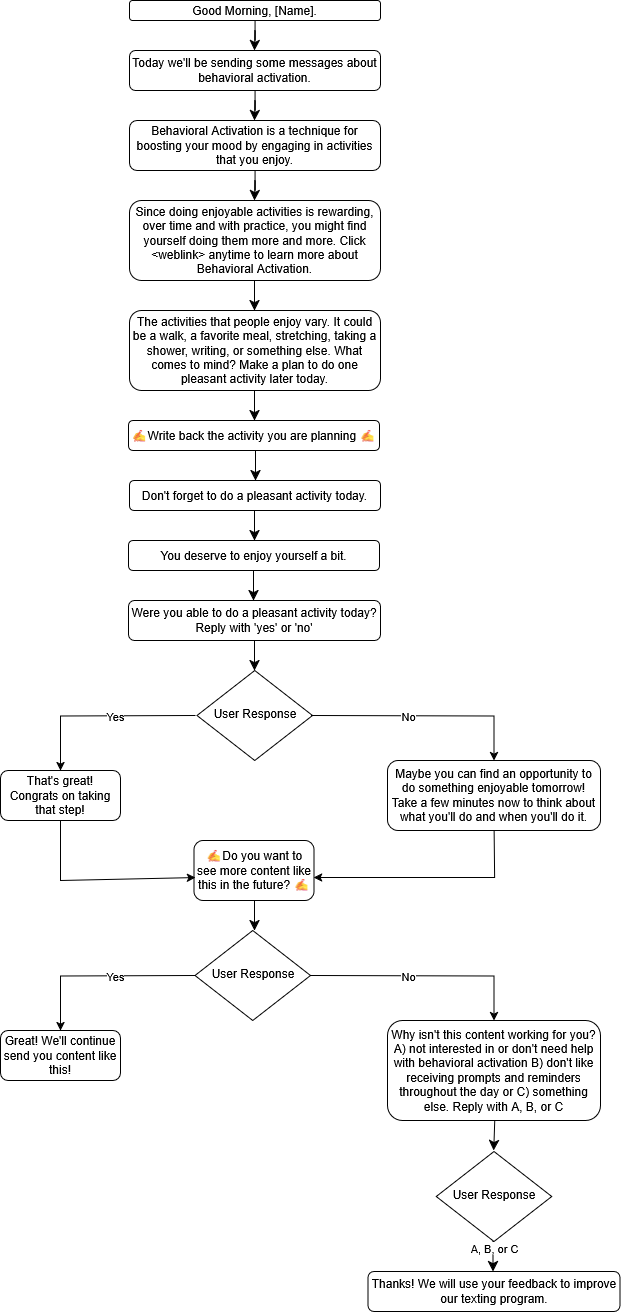}
        \caption{\topic{A message dialogue to communicate an evidence-based psychology principle}}
        \label{fig:pic2}
    \end{subfigure}
    \caption{\topic{Excerpts of text message dialogue flow}}
\end{figure}

\section{Hypothetical Scenarios Presented in  Workshops}
\label{app: scenarios}

\begin{table}[ht]
\caption{Summary of hypothetical scenarios presented in workshops}
\label{tab:scenario}
\begin{tabular}{|c|p{12cm}|}
\hline
\textbf{Number} & \textbf{Summary of the Scenario} \\
\hline
1 & Jenna is a single mother with multiple part-time employment responsibilities and a child with special needs. Her demanding schedule limits her ability to engage with text messages. \\
\hline
2 & Alex is a final-year university student overwhelmed by academic requirements, extra-curricular activities, and a part-time job. He finds the daily texts burdensome during such high-stress periods. \\
\hline
3 & Emma faces housing instability and is currently suffering from bronchitis. Her challenging life circumstances make it difficult to engage with the daily text messages. \\
\hline
\end{tabular}
\end{table}

\end{document}